\documentclass[twocolumn,showpacs,preprintnumbers,amsmath,aps,amssymb]{revtex4}
\usepackage{graphicx}% Include figure files
\usepackage{dcolumn}% Align table columns on decimal point
\usepackage{bm}% bold math

\begin{document}

%\preprint{Final Draft, June 2007}

\title{Noncommutivity and Scalar Field Cosmology }
\author{W. Guzm\'an$^1$}
 \email{wguzman@fisica.ugto.mx}
 \author{M. Sabido$^{1}$}
\email{msabido@fisica.ugto.mx}
 \altaffiliation[\\Member of  ]{Instituto Avanzado de Cosmolog\'ia (IAC).}%Lines break automatically or can be forced with \\
\author{J. Socorro$^{1,2}$}
\email{socorro@fisica.ugto.mx}
\affiliation{$^1$ Instituto de F\'{\i}sica de la Universidad de 
Guanajuato,\\
 A.P. E-143, C.P. 37150, Le\'on, Guanajuato, M\'exico\\
 $^2$ Facultad de Ciencias de la Universidad Aut\'onoma del Estado de 
M\'exico,\\
Instituto Literario No. 100, Toluca, C.P. 50000, Edo Mex, M\'exico.}%

\date{\today}% It is always \today, today,

\begin{abstract}
 In this work we extend and apply a previous proposal to study 
noncommutative cosmology to the FRW cosmological background coupled to 
 a scalar field, this is done in classical and quantum scenarios. In both  
cases noncommutativity is introduced in the gravitational field as well 
as 
 in the  scalar field through a deformation of  minisuperspace and  
are able to find  exact solutions. Finally, the effects of noncommutativity on the classical evolution are analyzed.
 \end{abstract}


\maketitle
%\section{Introduction}
It is a common issue in Cosmology  to make use of scalar
fields $\phi$ as the responsible agents of some of the most intriguing
aspects of our universe (see \cite{inflation} and references therein). Just to mention a few, we find that scalar
fields are used as the inflaton which seeds the
primordial perturbations for structure formation during an early
inflationary epoch; as the cold dark matter candidate responsible for
the formation of the actual cosmological structure and as the dark
energy component which seems to be driving the current accelerated
expansion of the universe. 
The key feature for such flexibility  of scalar fields
(spin-$0$ bosons) is the freedom one has to propose a \textit{scalar
  potential} $V(\phi)$, which encodes in itself the (non
gravitational) self-interactions among the scalar particles.

To study the very early universe one of the simplest approaches  
is Quantum Cosmology (QC). In this framework gravitational 
and matter variables have been reduced to a finite number of degrees of 
freedom. For homogenous cosmological models the metric 
depends only on time  and gives a model with a finite dimensional configuration 
space, {\it minisuperspace} \cite{ryan}. This simplification is used because a full quantum theory of
gravity does not exist (although two main candidates have been constructed: String 
Theory and Loop Quantum Gravity) 
 and  these approximate models  can be canonically quantized. 
 
Since the end of the XX century the old idea of noncommutative 
space-time \cite{snyder} has been rekindled, the renewed interest is a 
consequence  of the developments in M-Theory  and String 
Theory \cite{connes}. Although most of the work of noncommutative 
field theory has been done in connection with Yang-Mills theory \cite{Douglas:2001ba}, 
several noncommutative models of gravity have been constructed 
\cite{chandia}. All of these formulations of gravity in noncommutative 
space-time share a common issue; they  are highly nonlinear 
and solutions to the noncommutative equations of motion are incredibly 
difficult. One may expect that at very early times non trivial 
effects of noncommutativity  could affect the evolution of the 
universe, making the study of noncommutative cosmological models 
an interesting testing ground for noncommutativity. Unfortunately, as already mentioned, the 
complexity of  the field equations for noncommutative gravity makes 
a direct calculation of cosmological models a very complex task.
On this line of reasoning, in the last few years there have been several attempts to study the 
possible effects of noncommutativity in the cosmological  and inflationary scenario
\cite{ncqc, bran,barbosa1}. In  \cite{ncqc} the authors in a cunning way 
avoid the difficult technicalities of analyzing noncommutative 
cosmological models when these are derived from the full 
noncommutative theory of gravity. They achieved this by introducing the effects 
of noncommutativity in quantum cosmology by  deforming the  
minisuperspace variables and introducing the Moyal product of functions in the 
Wheeler-DeWitt equation (WDW), similar to noncommutative 
quantum mechanics \cite{jabbari}. Other authors have analyzed some phenomenological effects of noncommutativity in cosmology, \cite{bran} but in their analysis they neglect the noncommutative effects on the gravitational sector.
The aim of this paper is to construct a noncommutative formulation of a simple scalar field cosmological model in which the matter and gravitational sector are affected by the noncommutative deformation. 

Let us start with the line element for a homogeneous, flat 
and isotropic
universe, the  Friedmann-Robertson-Walker (FRW) metric
\begin{equation}
ds^2= -N^2(t) dt^2 + e^{2\alpha(t)}\left[dr^2
+r^2 d\Omega^2 \right] , 
\label{frw}
\end{equation}
where $ a(t)=e^{\alpha(t)}$ is the scale factor and $\rm N(t)$ is the 
lapse function. The classical evolution is obtained by 
solving Einstein's field equations. Using the energy momentum tensor of 
a scalar field we get the
Einsten-Klein-Gordon equations
\begin{equation}
H^2=\rm \frac{8\pi G}{3}\left 
(\frac{1}{2}\dot{\phi}^2+V(\phi)\right ),~ ~\ddot{\phi}+3H\dot{\phi}= -\frac{\partial 
V(\phi)}{\partial\phi},
\label{fried}
\end{equation}
where $ H=\dot{a}/a$,  $ 
\rho=\frac{1}{2}\dot{\phi}^2+V(\phi)$ and 
$ p=\frac{1}{2}\dot{\phi}^2-V(\phi)$. As stated at the beginning, by 
choosing appropriate  {potentials}, different aspects of the universe can be modeled.

For our purposes the Hamiltonian formulation is more appropriate and is constructed from the effective 
action 
\begin{equation}
 S =\int  dx^4 \sqrt{-g} \left[ R + 
 \frac{1}{2}g^{\mu \nu}\partial_\mu \phi \partial_\nu \phi+ 
 V(\phi)
  \right] \, ,
\label{accion}
\end{equation}
where we have used the units $8\pi G=1$.
Instead of solving Einstein's equations for the FRW metric with the scalar field
energy momentum tensor, we will be using  the Hamiltonian dynamics derived from Eq.(\ref{accion})  as well as one of the most used and better understood scalar 
field potentials, the exponential  potential 
 $\rm V(\phi) = V_0 e^{-\lambda\phi}$. This potential has the advantage that analytical solutions can be found and for appropriate values of the parameters inflation can be obtained. Unfortunately, when used in connection with inflation there is no mechanism to end the inflationary epoch.
 We can write the  canonical Hamiltonian by means of the Legendre 
transformation 
and arrive to the FRW Hamiltonian
\begin{equation}
\rm {\cal H}=\frac{N}{12} e^{-3\alpha} \left[
\Pi_\alpha^2 - 6\Pi_\phi^2 - 12  e^{6 \alpha}  V(\phi)
\right]. \label{uno}
\end{equation} 
The Poisson algebra for the minisuperspace variables is 
$\{\alpha,\phi\}=0,~ \{\Pi_{\alpha},~\Pi_{\phi}\}=0,~ 
\{\alpha,\Pi_{\alpha}\}=1,~\{\phi,\Pi_{\phi}\}=1$, if we set $N=1$ we recover the equations 
derived from GR. To simplify the calculations  we will be working in 
the gauge $N=12e^{3\alpha}$, the advantage is that in this gauge we can find exact analytical solutions to the noncommutative  quantum 
models. Because of the reparametrization  invariance of the theory  the physical implications are independent of the gauge.
For $N=12e^{3\alpha}$ the equations of motion take the simple form
\begin{eqnarray}
\rm \dot{\alpha}&=&\rm 2 \Pi_\alpha,\qquad 
\dot{\Pi}_\alpha=72e^{6\alpha}V(\phi),\nonumber\\
\rm \dot{\phi}&=&\rm -12\Pi_\phi,~~ 
\dot{\Pi}_\phi=12e^{6\alpha}\frac{dV(\phi)}{d\phi}.
\label{classical2}
\end{eqnarray} 
As already pointed, we will be using the exponential potential $
V(\phi)=e^{-\lambda\phi}$. Eq.(\ref{classical2}) is simplified by making the canonical transformation
\begin{eqnarray}
\rm x&=&\rm -6\alpha+\lambda\phi,\qquad~~ 
\rm \Pi_x= \frac{1}{\lambda^2-6} \left(\Pi_\alpha +\lambda\Pi_\phi\right), \label{cano}\\
\rm y&=&-\sqrt{6}\lambda\alpha+\sqrt{6}\phi,
~~
\Pi_{\rm y}=\frac{1}{\sqrt{6}(6-\lambda^2)} \left(\lambda\Pi_\alpha 
+6\Pi_\phi\right).\nonumber
\label{trans}
\end{eqnarray}
In this new set of minisuperspace variables we will be making our 
analysis, 
the classical Hamiltonian Eq.(\ref{uno}) in the new variables 
has the simple form
\begin{equation}
\rm {\cal 
H}=-\beta \Pi^2_x+\beta \Pi^2_y-12V_0e^{-x} \approx 0,
\label{uno-variables}
\end{equation} 
where $\beta$ is defined as $\beta\equiv6(\lambda^2-6)$. In these variables the equations of motion are considerably 
simplified, in particular $\dot{ \Pi}_{\rm y}=0$ gives a constant of motion, which helps to solve the rest of the equations, actually the canonical transformation was used in order to have a simplified Hamiltonian that has at least one constant of motion.
The solutions in the minisuperspace variables  $\rm (x,y)$ are
\begin{equation}
 \rm x(t)=\rm \ln \left(\cosh^2\sqrt{12\beta 
V_0}~t\right),~
y(t)=\rm 2\beta \Pi_{y_0}(t-t_0).
\end{equation}
After applying the inverse canonical 
transformation we get the solutions in the original variables
\begin{eqnarray}
 \phi(t)&=&\rm \frac{1}{6-\lambda^2}\left(2\sqrt{6}\beta 
\Pi_{y_0}t-\lambda\ln\left(\cosh^2\sqrt{12\beta V_0}~t\right) 
\right),\nonumber \\
 \alpha(t)&=&\rm \ln \left(\cosh^{-\frac{2}{6-\lambda^2}}\left[ 
\sqrt{12\beta V_0}~t\right] \right)-2\lambda\sqrt{6}\,\Pi_{y_0} t,
\end{eqnarray}
and can easily calculate the expression for the scale factor $a(t)$. In these solutions the parameter $t$ is not the usual time that is used in 
cosmology. In order to have exactly the same solutions, we need to work in the gauge $N=1$ and
to  find the inflationary epoch we apply the slow roll condition  $\epsilon=\left(V'(\phi)/V(\phi)\right)^2<<1$, this yields the condition $\lambda<\sqrt{2}$. Carrying out the same analysis in the gauge and variables we have chosen, we obtain the same result, this is to be expected as the theory is invariant under time reparametrization.

The original proposal for noncommutative cosmology, was done at the quantum level \cite{ncqc}. The simplest approach for quantum cosmology is  canonical quantization and is achieved by applying the well-known
result $\rm {\cal H}=0$.
The WDW equation for this model is achieved by the usual
identifications, $ \Pi_{\rm x}$=$ -i \partial_{\rm x}$ and $ \Pi_{\rm y}$=$ -i \partial_{\rm y}$
 in Eq.~(\ref{uno-variables}) and gives a Klein-Gordon type equation
 \begin{equation}
\rm  \frac{\partial^2 \Psi}{\partial x^2}
 - \frac{\partial^2 \Psi}{\partial y^2} + \frac{2 V_0}{6-\lambda^2 }
 e^{-x} \Psi=0.
\label{modified}
\end{equation}
In this formalism, $\Psi$ is called the wave function of the universe. To extract a normalizable wave function we need to construct wave 
packets to form a Gaussian state from which some physical information can be obtained.
The proposal to introduce the noncommutative minisuperspace deformation  is achieved by introducing the following commutation relation between the minisuperspace variables
\begin{equation}
[\rm x,\rm y]=i\theta,
\label{ncms}
\end{equation}
this can be seen as an effective noncommutativity that could arise from a fundamental noncommutative theory of gravity. For example, if we start with the Lagrangian derived in \cite{chandia} (this is a higher order Lagrangian and is expanded in the usual noncommutative parameter), the noncommutative fields are a consequence of noncommutativity among the coordinates and then the minisuperspace variables would inherit  some effective noncommutativity. This we assume to be encoded in (\ref{ncms}), otherwise we would have a very complicated Hamiltonian for the higher order Lagrangian.

This effective noncommutativity
can be formulated in terms of product of functions of the minisuperspace variables, with the Moyal star product of functions. The noncommutative WDW (NCWDW) equation is obtained by replacing  the products of functions by  star products.
We can show \cite{jabbari} that the effects of the Moyal star product are reflected only in a shift in the potential $V(\rm x,\rm y)\star\Psi(\rm x,\rm y)=V(\rm x+\frac{\theta}{2}\Pi_{\rm }y,\rm y-\frac{\theta}{2}\Pi_{\rm x})\Psi(x,y)$. 
Applying this to Eq. (\ref{modified}) we can write
\begin{equation}
\rm  \left(-\frac{\partial^2 }{\partial x^2}
 + \frac{\partial^2}{\partial y^2} -\gamma
 e^{-\left (x-\frac{i}{2}\theta\frac{\partial}{\partial y}\right 
)}\right) \Psi=0,
\end{equation}
and the solutions are given by
\begin{equation}
\rm \Psi^{NC}_\eta(x,y)=\rm e^{\pm i\eta {\rm y}}\left [J_{\pm i2\eta}\left( 
\sqrt{\vert\gamma\vert}e^{ -\frac{x}{2}-\frac{\eta\theta}{4}}\right) 
\right ].
\end{equation}
These  solutions give the quantum description of the noncommutative universe. The wave packet can be constructed as in the commutative case, but unfortunately  is difficult to obtain physical information. In \cite{ncqc} the authors arrive to a similar equation, but they are interested only in the quantum epoch of the Kantowski-Sachs universe. By solving the NCWDW equation and plotting the probability density after constructing a wave packet.
In order to extract physical information we may try to find the classical evolution for this model, this can be achieved by several equivalent ways (i.e. WKB type analysis, Hamiltonian dynamics, etc.). 

For the classical evolution, we start by proposing 
that the noncommutative nature of the minisuperspace variables can be encoded in a deformation of the Poisson structure this has the advantage that in this approach the Hamiltonian is not modified. For our model the new algebra for the 
minisuperspace variables is, $\{\rm x,\rm y\}= \theta$, $\{\Pi_{\rm x},~\Pi_{\rm y}\}=0$, 
$ \{ \rm x,\Pi_{\rm x} \}=1$, $ \{\rm y,\Pi_{\rm y} \}=1$.
From this algebra we find the noncommutative equations of motion 
\begin{eqnarray}
 \dot{\rm x}&=&-2\beta \Pi_{\rm x},~~~~~~~~~~~~~~~~
\dot{\Pi}_{\rm x}=-12V_0e^{-\rm x},\nonumber\\
 \dot{\rm y}&=&\rm 2\beta \Pi_{\rm y}-12\theta V_0e^{-\rm x},~~ \dot{\Pi}_{\rm y}=0,
 \label{nceqm}
\end{eqnarray}
as in the commutative model the equation associated to the momenta 
$\Pi_{\rm y}$ is a constant of motion, this simplifies the calculations 
considerably. Solving this set of equations we arrive to the classical 
evolution for noncommutative cosmology
\begin{eqnarray}
\rm x(t)&=&\rm \ln\left[\cosh^2\left (\sqrt{12\beta 
V_0}t\right)\right],\label{nc}\\
\rm y(t)&=&\rm 2\beta \Pi_{y_0}t 
+\theta\left(\frac{12V_0}{\beta}\right)^{\frac{1}{2}}\textrm{tanh}\left(\sqrt{12\beta 
V_0}t\right).\nonumber
\end{eqnarray}
From the canonical transformation, Eq.(\ref{trans}), we reconstruct the 
noncommutative scale factor
and the time dependence of scalar field. As in the commutative case, the parameter $t$ does not correspond to the usual cosmic time and as already stressed, the physical implications are independent of the chosen gauge.

We can see in Eq.(\ref{nc}) that the noncommutative deformations only modify the variable $\rm y$. Furthermore, the hyperbolic tangent asymptotically goes to $\pm 1$, then the effects of noncommutativity in the evolution of the universe can only be felt for a short time, after which the behavior is exactly the same as in the commutative case, non the less there is the possibility that the effects of this minisuperspace noncommutativity may change the dynamics during this short period of time. In order to study this possibility we will analyze the inflationary scenario that can be constructed from this noncommutative model. For the commutative case, the potential that we are using gives inflation if we fix the parameter in the potential to $\lambda<\sqrt{2}$. One of the advantages of this inflationary potential, is that exact analytical solutions can be found, but  the lack of a mechanism to end inflation is its biggest drawback. This can be seen from the constant value we get for the slow roll parameter $\epsilon=\lambda^2/2$, eliminating any possibility to violate the slow roll condition $\epsilon<1$. To study some consequences of the noncommutative model we start by calculating $\epsilon$, this can easily be done from the definition of the slow roll parameter and the noncommutative solutions. To first order in $\theta$, we find
\begin{equation}
\epsilon=\frac{\lambda^2}{2}\left[1-\theta\mu\left(1- \frac{\sqrt{6}}{\lambda}\tanh{[3\mu(\lambda^2-6)t]}\right)\right],
\end{equation}
where $\mu=\sqrt{\frac{8V_0}{\lambda^2-6}}$. We can write the slow roll parameter as the sum of to contributions $\epsilon=\epsilon_c+\epsilon_{nc}$, the first term corresponds to the usual slow roll parameter and the second one to the noncommutative contibution. As expected in the limit $\theta\to 0$ we get the usual condition for inflation $\lambda<\sqrt{2}$, also for $t\to\infty$, the slow roll parameter tends to a constant value, meaning that the effects of the noncommutative deformation can not be seen.
If we  start in an inflationary epoch for the commutative case ($\lambda<\sqrt{2}$), in the noncommutative case the slow roll parameter also starts at a value smaller than 1, so in both cases we have inflation, but as the universe evolves the value of $\epsilon$ in the noncommutative model increases and by choosing appropriately the parameters of the potential, as well as the value of $\theta$, we can violate the slow roll condition and end inflation; this is in contrast with the commutative case. Then we can argue that the introduction of the noncommutative minisuperspace gives  a new mechanism to end inflation. Also the time that  takes to end inflation is regulated by the value of $3\mu(\lambda^2-6)$, due to the freedom that we have on the parameters of the model this time  can be fixed to give the correct number of e-folds.
\begin{figure}
\includegraphics[width=6cm]{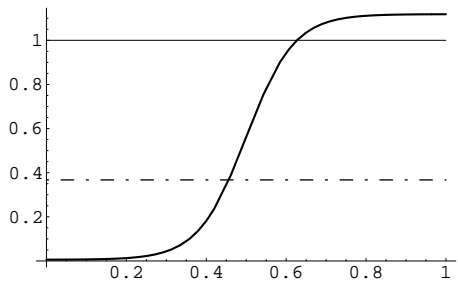}
\caption{\label{fig:inflation}
The dotted line corresponds to the commutative slow roll parameter $\epsilon_c$, the thin line is the value of $\epsilon$ that breaks inflation, the thick line corresponds to the new slow roll parameter. }
\end{figure}
In Fig.(\ref{fig:inflation}), we have plotted the slow roll parameter $\epsilon$ for the commutative and noncommutative models. The values we have taken are $\lambda=6/7$, $\theta=1$ and $\mu=0.53$, we can see that before $t=0.6$, $\epsilon <1 $, after that time the slow roll condition is violated and inflation ends, after which the value of $\epsilon$ is asymptotically constant.

\begin{figure}
\includegraphics[width=6cm]{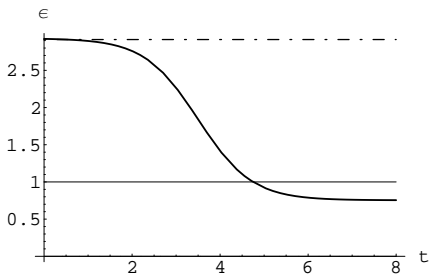}
\caption{\label{fig:de} 
The dotted line corresponds to the commutative slow roll parameter $\epsilon_c>1$ so there is no acceleration, the thin line corresponds to $\epsilon=1$. In the noncommutative model (thick line), there is a transition from a non accelerating universe to an accelerating one.}
\end{figure}

The other interesting possibility is in connection with the current acceleration of the universe. If we analyze the same model but start with a value of $\lambda>\sqrt{2}$, again we can fix the values of the parameters in order to have a transition from a non accelerating universe to an accelerating one, this can be seen from the behavior of the slow roll parameter which goes from $\epsilon>1$ to $\epsilon<1$ and again the parameters can be tuned to have the correct value for the current observed acceleration. In Fig.(\ref{fig:de}), we have taken for the parameters the values $\lambda=\sqrt{2}+1$, $\theta=.23$ and $\mu=1.6$. 
By analyzing the time evolution of  the noncommutative slow roll parameter we can see  that we have a transition from a non accelerating universe to an accelerating one, this behavior is completely different to the commutative case.

Summarizing,  we have introduced a noncommutative deformation to the minisuperspace of scalar field cosmology in the gravitational and matter sectors of the model,
this has been done at the quantum and classical levels. The corrections of the noncommutative deformation have been analyzed  at the classical level and the most striking  effect, is the presence of a mechanism that ends inflation.
This mechanism also permits that a non accelerating universe after a period of time can start an acceleration. This might  give some evidence on a possible relationship between dark energy an noncommutativity, although this is only an speculative statement further research in this direction might establish or discard this possibility.
 As a final remark, we have shown in this simple toy model that noncommutativity may have very important implications in the evolution of the universe.

\acknowledgments{This work was partially supported by CONACYT grants  
47641, 51306-F
  and PROMEP grants UGTO-CA-3, UGTO-PTC-085, M.S. is also supported by CONCYTEG grant 0716K662062. }

\end{document}